\documentclass[aps,prl,twocolumn,floatfix,showpacs]{revtex4}
\usepackage{graphicx}
\usepackage{amssymb}
\usepackage{amsmath}

\newcommand{\ssm}{\scriptscriptstyle\rm}
\renewcommand{\phi}{\varphi}

\begin{document}

\title{Dynamically Generated Double Occupancy as a Probe of Cold Atom Systems}

\author{S.D. Huber}
\affiliation{Theoretische Physik, ETH Zurich, CH-8093
Z\"urich, Switzerland}
\author{A. R\"uegg}
\affiliation{Theoretische Physik, ETH Zurich, CH-8093
Z\"urich, Switzerland}

\date{\today}

\begin{abstract}
The experimental investigation of quantum phases in optical lattice systems
provides major challenges. Recently, dynamical generation of double
occupancy via modulation of the hopping amplitude $t$ has been used to
characterize the strongly correlated phase of fermionic atoms. Here, we
want to validate this experimental technique with a theoretical study of
the driven Hubbard model using analytic methods.  We find that conclusive
evidence for a Mott phase can be inferred from such a measurement, provided
that sufficiently low temperatures $k_{\ssm B}T \ll t$ can be reached.
\end{abstract}


\maketitle

\begin{figure}[b]
\includegraphics{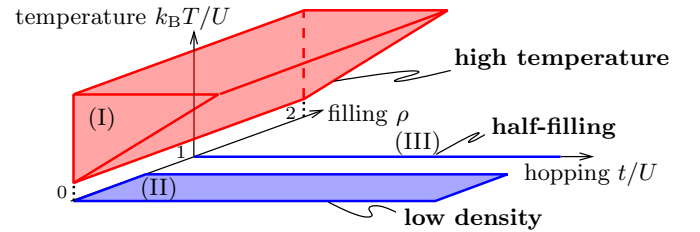}
\caption{
(color online) Regimes where the dynamically generated double occupancy probe
is analyzed: High temperatures (I) are discussed in the atomic limit. The
regime at zero temperature and low densities (II) involves the solution of a 
two-particle problem; the situation at half filling and $T=0$ (III) is done within a
slave-spin mean-field analysis.
}
\label{fig:regimes}
\end{figure}

The recent progress in cooling and manipulating cold fermionic gases
in optical lattices allows us to investigate phenomena at ever lower
temperatures, where intriguing many-body effects generate new (quantum)
phases which are of much interest in condensed matter systems as well.
While the high tunability of atoms in optical lattices is a landmark
advantage of these systems, the availability of methods allowing for their
proper characterization lags behind the sophistication reached in solid
state physics. This lack in available experimental probes gains importance,
now that an increasingly larger class of non-trivial quantum phases are
in reach.  Their definitive identification requires the translation of
the defining properties, usually expressed in terms of thermodynamic and
transport coefficients, to the probes available for cold atomic systems.
Recently two experiments addressed the question of strongly correlated
fermions in an optical lattice \cite{Jordens08,Schneider08}. Schneider
{\it et al.} \cite{Schneider08} concentrate on the compressibility
\cite{Scarola08}, while J{\"o}rdens {\it et al.} \cite{Jordens08} have
made use of a new probe, the dynamically generated double occupancy
(DGDO) proposed by  Kollath {\it et~al.} \cite{Kollath06a}, in order to
identify a Mott insulator state of strongly repulsive fermionic $^{40}$K
atoms. This technique, where the signal is the change in double occupancy
after dynamical modulation of the hopping amplitude $t$, represents a prime
example of a measurement without analog in classical solid state physics
\cite{Folling06}. Using different analytic methods in separate regions
of the parameter space, cf.  Fig.~\ref{fig:regimes}, this Letter aims at
a theoretical characterization of the DGDO scheme and its suitability to
describe the transition or crossover to the fermionic Mott-insulator.

Ultracold fermions in an optical lattice are well described by the
Hubbard model
\begin{equation}
\label{eqn:hubbard}
H \!=\! 
-t \sum_{\langle i,j\rangle, \sigma}(c_{i\sigma}^{\dag}c_{j\sigma}^{} 
  + {\rm h.c.})+ U \sum_{i} D_{i} \!=\!-t K + U H_{\ssm int}, 
\end{equation}
where the spin degrees of freedom are realized by two hyperfine
states.  Here, $c_{i\sigma}^{\dag}$ are fermionic operators
creating particles with spin $\sigma$ in a Wannier state at site $i$,
$D_{i}=n_{i\uparrow}n_{i\downarrow}$ measures the double occupancy at site
$i$, and the sum $\langle i,j\rangle$ runs over nearest neighbors for all
$N$ lattice sites. Although the phase diagram for the three dimensional
system is not precisely known yet, the qualitative form was determined
(c.f. \cite{Imada98,De-Leo08} and references therein). In particular,
at half-filling on a cubic lattice, an anti-ferromagnetic phase below $T
\lesssim T_{\mbox{\tiny N\'eel}}\leq 0.3\, T_{\ssm F}$ is expected; whereas
in strongly frustrated systems, typical estimates limit the temperature for
(paramagnetic) Mott physics to a regime below $0.1\,T_{\ssm F}$ ($T_{\ssm
F}$ denotes the Fermi temperature).

The two defining properties of a fermionic Mott insulator are the
interaction-induced incompressibility and a vanishing Drude-weight in
the optical conductivity  \cite{Imada98}. While the first is related to
a gap $\Delta \mu$ in the charge spectrum and can be used as a probe in
the cold atom setup as well, the measurement of the second one, relying on
transport properties, is more difficult to perform.  From this viewpoint,
the implementation of the Hubbard Hamiltonian with cold atoms not only
allows us to study this unsolved model in a controlled environment but also
poses inherently new questions.  In comparing theoretical and experimental
results, it is of interest to know which features in the calculation (which
is based on a uniform system) survive the translation to the (inhomogeneous)
experimental setup.  We thus characterize the DGDO technique in a wide
parameter regime and conclude that frequency-integrated and bond-averaged
quantities capture the desired information with the least obstructions
due to confinement-induced changes of line shapes.

\begin{figure}[b]
\includegraphics{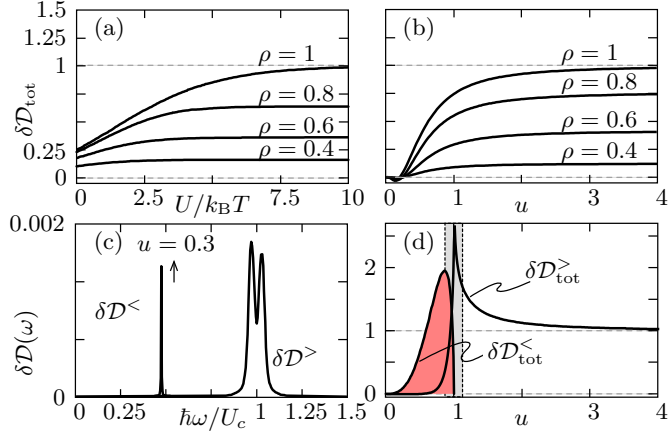}
\caption{
(color online) Characteristic DGDO signal for
the different regimes (I)-(III) of Fig.~\ref{fig:regimes}. (a) For the high
temperature region (I), we find an increase in $\delta{\mathcal D}_{\ssm tot}$
for increasing $U/T$ and a saturation value $\delta{\mathcal
D}_{\ssm tot} \rightarrow \rho^{2}$ at $U\gg T$. (b) For the low
density phase at $T=0$, we find the same behavior with a saturating response
at $u=U/U_{c} \gg 1$ where $\delta{\mathcal D}_{\ssm tot} \to
\rho^{2}[1-\sin(4\pi\rho)/4\pi\rho]$. The response of the system at
half-filling and $T=0$ (III) changes its behavior significantly at $u=1$. (c)
Spectral DGDO signal $\delta{\mathcal D}(\omega)$ with peaks at $\hbar\omega
=\Delta$ and around $\hbar\omega\approx \max(U_{c},U)$ identifying
transitions between the Gutzwiller and upper Hubbard- and those between the
lower and higher Hubbard bands, respectively. (d) Total weights
$\delta{\mathcal D}^{<,>}_{\ssm tot}$ accumulated in the two peaks in (c).
Shown in grey is the fluctuation region $|u-1|<u^{\ssm fl}$ where our
mean-field analysis is not valid.
}
\label{fig:dido-tot}
\end{figure}

Our theoretical analysis of the DGDO technique involves exactly solvable
limiting cases of the Hamiltonian (\ref{eqn:hubbard}), namely, the atomic
limit ($t=0$) and the two-particle problem, see Fig.\ \ref{fig:regimes},
with results relevant in the regimes $U,T\gg t$, as well as for filling
$\rho\ll 1$ (we set $k_{\ssm B} = 1$). On the other hand, in order to
discuss the Mott physics within the DGDO technique at low temperatures,
cf.~Fig.~\ref{fig:regimes}(III), we introduce and apply a slave-spin
mean-field approximation which captures the most relevant physics on both
the $t$ and $U$ scale. Our results for the DGDO signal are summarized in
Fig.~\ref{fig:dido-tot}.

We first derive the expression for the second-order response function
describing the DGDO $\delta\mathcal{D}$; exciting the system during the
time $\Delta\tau_{\ssm mod}$ through a (isotropic) modulation $H_{\ssm
mod}=\delta t \cos(\omega \tau) K$ of the hopping amplitude, the task is
to calculate the expectation value
\begin{equation} \label{eqn:d-tau} \delta D(\tau) = \langle \psi(\tau) |
\sum_{i} D_{i} | \psi(\tau) \rangle- \langle 0 |  \sum_{i} D_{i} |0\rangle,
\end{equation}
of the double-occupancy operator, where $|0\rangle$ denotes the unperturbed
ground state. We write the perturbed wave function $\psi(\tau)$ in the
interaction representation, $|\psi(\tau)\rangle=\exp(iH\tau/\hbar)
{\rm T_{\tau}} \exp[-(i/\hbar)\int_{-\infty}^{\tau}d\tau' H_{\ssm
mod}(\tau')] |0\rangle$, where ${\rm T}_{\tau} $ denotes time
ordering.  Expanding (\ref{eqn:d-tau}) to second order in the hopping
modulation $\delta t$ and keeping only non-oscillatory terms $\delta
D(\tau)=\pi \delta t^{2}\delta{\mathcal D}(\omega)\Delta\tau_{\ssm
mod}/2\hbar^{2}+\mathrm{osc.}+O(\delta t^{3})$, we arrive at the spectral
transition rate (the time-averaged first-order response vanishes)
\begin{equation}
\delta {\mathcal D}(\omega) = \sum_{n} \langle n| \delta D  |n\rangle
|\langle n | K | 0 \rangle|^{2} \delta(\omega-\omega_{n0}).
\label{eq:Domega}
\end{equation}
Here, $|n\rangle$ denote the excited states with energy $\hbar\omega_{n0}=
\hbar\omega_{n}-\hbar\omega_{0}$. Note, that in the current experiment
\cite{Jordens08}, $\delta D(\tau)$ is not $\propto \Delta\tau_{\ssm mod}$,
however  \cite{note:boston2}. We define the integrated quantity
\begin{equation}
\label{eqn:dido-tot}
   \delta {\mathcal D}_{\ssm tot}
  = \frac{2}{Nz}\int d\omega\, \delta{\mathcal D}(\omega),
\end{equation}
normalized to the number $Nz/2$ of bonds in the system ($z = $ coordination
number). In the non-interacting limit ($U=0$), the eigenstates of
(\ref{eqn:hubbard}) are those of $K$ and no excitations are induced, hence
$\delta {\mathcal D}_{\ssm tot}$ vanishes and a finite interaction $U$
is required to generate a finite result.  In the following, we evaluate
(\ref{eqn:dido-tot}) in the regimes (I)--(III).

In the atomic limit ($t=0$, region I), the calculation of the {\it change}
in the double occupancy reduces to a two-site problem and $\delta {\mathcal
D}_{\ssm tot}$ is given by the square of the probability to find a singly
occupied lattice site; its rapid increase signals the crossover from weak to
strong correlations. In the limit $U\gg T$, $\delta {\mathcal D}_{\ssm tot}$
saturates at $\rho^2$ for $\rho\leq1$ and at $(2-\rho)^2$ for $\rho\geq 1$
[cf.\ Fig.~\ref{fig:dido-tot}(a)]. The atomic limit, valid in the regime
$U,T\gg t$, offers a good starting point for analyzing current experiments
\cite{Jordens08,Scarola08}.

In the low density limit ($\rho<1$, region II), we analyze the DGDO
by exactly solving the two-particle problem with Hubbard interaction
\cite{Maldague77, Winkler06}.  One then expects a signature at finite
frequency of order $U$, as two fermions (in a singlet state) on the same
site cost an energy $U$, irrespective of the density. We only consider
the $d=1$-dimensional case here, the extension to higher dimensions is
straightforward. The two-particle wave function in the singlet channel,
$\psi_{\ssm scat}(r)=\langle r | k_{1},k_{2}\rangle_{\ssm scat}$, with
$r$ the relative coordinate and $k_{1}$ and $k_{2}$ the momenta of the two
atoms, can be obtained by solving the Lippmann-Schwinger equation (we ignore
the center of mass motion). Expanding the solution $\psi_{\ssm scat}(r)$
in Bloch waves $\psi_{\ssm Bloch}(r)=\langle r | k_{1},k_{2}\rangle_{\ssm
Bloch}$ we find the amplitudes
\begin{equation}
_{\ssm Bloch}\langle q,k' | q,k \rangle_{\ssm scat} = 
\delta_{k,k'}+
\frac{(U/N) f(q,k)}{\epsilon_{q,k}-\epsilon_{q,k'}+i\eta},
\end{equation}
with $\epsilon_{k,q}=4t\cos(q)\cos(k/2)$ the bare dispersion and
$f(q,k)=[1+U/4ti|\sin(k/2)\cos(q)|]^{-1}$ the scattering amplitude.  Here,
$q=(k_{1}+k_{2})/2$ and $k=k_{1}-k_{2}$ are the total and relative momenta of
the two atoms. In addition, a bound state $\psi_{\ssm bs}(r)$ with amplitudes
\begin{equation}
_{\ssm Bloch}\langle q,k' | q\rangle_{\ssm bs} = 
(2U^{3}/\epsilon^{\ssm bs}_{q})^{1/2}/(\epsilon_{q,k'}-\epsilon^{\ssm bs}_{q})
\end{equation}
and energy $\epsilon^{\ssm bs}_{q}=\sqrt{U^{2}+[4t\cos (q)]^{2}}$ is found.
Rewriting $K= (H_{\ssm int}-H)/t$ with $H_{\ssm int}=U\delta_{r,0}$, we find
for the total DGDO (\ref{eqn:dido-tot}) per singlet pair $\{k_{1},k_{2}\}$,
\begin{equation}
\delta {\mathcal D}_{\ssm tot}^{k_{1},k_{2}}=
[\psi_{\ssm bs}^{2}(0)\!-\!\psi_{\ssm scat}^{2}(0)]
(U/t)^{2}
\psi_{\ssm bs}^{2}(0)\psi_{\ssm scat}^{2}(0);
\end{equation}
the above result originates from process where two particles in the
scattering state $|q,k\rangle_{\ssm scat}$ combine to a repulsively
bound pair $|q\rangle_{\ssm bs}$. To obtain the full result at small
but finite density $\rho$, we have to calculate $\delta{\mathcal
D}_{\ssm tot}=\sum_{k_{1},k_{2}} n_{k_{1},k_{2}} \delta {\mathcal
D}_{\ssm tot}^{k_{1},k_{2}}$, where $n_{k_{1},k_{2}}$ is the ground
state expectation value to find a pair $\{k_{1}, k_{2}\}$ in the
singlet channel, see Ref.~\cite{Maldague77}.  The final result, cf.\
Fig.~\ref{fig:dido-tot}(b), shows again an increasing signal for small
$U/t$ and a saturation at large $U/t$, with $\delta {\mathcal D}_{\ssm
tot}\rightarrow \rho^{2}[1-\sin(4\pi\rho)/4\pi\rho]$. While the derivation
assumes low density, the result up to $\rho=1$ are shown for comparison
with the other approaches \cite{Maldague77}.

We now turn to the low-temperature regime of the half-filled Hubbard
model (\ref{eqn:hubbard}) in three dimensions, the region III in Fig.\
\ref{fig:regimes}. Depending on the amount of Fermi-surface nesting,
an anti-ferromagnetic phase intervenes with the Mott physics. For the
optical lattice implementation this would require an initial cooling to
about $T_i\approx 0.02\, T_{\ssm F}$ \cite{De-Leo08}.  Within the slave-spin
mean-field treatment used here, the possibility for the appearance of such
a magnetic order is ignored. All our results below can be obtained within
the four-boson approach of Kotliar and Ruckenstein \cite{kotliar86} which
builds on earlier slave-boson methods \cite{Barnes76,Coleman84}. However,
we apply a minimal formalism reminiscent of the slave-spin \cite{Medici05a}
or the slave-rotor formulation \cite{Florens04} which captures all aspects
relevant to our present discussion.

On each lattice site, we introduce an auxiliary pseudospin-1/2 ${\bf
S}$ with eigenstates (of $S_z$) $|+\rangle$ ($|-\rangle$) encoding
double and empty (singly) occupied sites with excitation energies of
order $U$. In addition, auxiliary fermionic operators $f_{\sigma}$ are
introduced to describe the low-energy quasiparticle degrees of freedom.
The physical creation (annihilation) operators of the original model then
are represented as $c_{\sigma}^{(\dag)}=2S^xf_{\sigma}^{(\dag)}$. The
physical states in the enlarged Hilbert-space are $|e\rangle =|+,0\rangle$,
$|d \rangle=f_{\uparrow}^{\dag}f_{\downarrow}^{\dag}|+,0\rangle$,
$|\!\!\!\uparrow\rangle=f_{\uparrow}^{\dag}|-,0\rangle$, and
$|\!\downarrow\rangle=f_{\downarrow}^{\dag}|-,0\rangle$, where $|0\rangle$
is the vacuum of the $f$-fermions. The projection onto the physical
subspace is achieved by imposing for each site $i$ the constraint
$S_i^z+1/2=(\sum_{\sigma}f_{i\sigma}^{\dag}f_{i\sigma}^{}-1)^2$; as a
result, $H_{\ssm int}$ involves solely pseudospin operators $S_i^z$.

Within our mean-field solution, we assume product states in pseudo spin and
fermion degrees of freedom, thereby relaxing the above constraint. Within
this approximation, the canonical anti-commutation relations are preserved
on average, $\langle\{c_{i\sigma}^{},c_{j\sigma'}^{\dag}\} \rangle=4\langle
S_i^xS_j^x\rangle\langle\{f_{i\sigma}^{\phantom{\dag}}, f_{j\sigma'}^{\dag}
\}\rangle=\delta_{\sigma\sigma'}\delta_{ij}$, where $\langle \dots \rangle$
denotes the average over mean-field eigenstates. A a consequence, the
single-particle spectral weight is preserved as long as the spin identity
$(S_{i}^{x})^{2}\equiv 1/4$ is respected.

We obtain two effective mean-field Hamiltonians: The fermion problem assumes
the form of a non-interacting tight-binding Hamiltonian with a hopping
amplitude renormalized by a factor $g_{ij}=4\langle S_i^xS_j^x\rangle$,
with $i,j$ nearest neighbors. On the other hand, the pseudospin problem
reduces to the transverse Ising model
\begin{equation} 
   H_{\mathrm{\ssm TIM}}=-\sum_{\langle i,j\rangle}J_{ij}S_i^xS_j^x
   +h\sum_iS_i^z, 
   \label{eq:HTIM} 
\end{equation}
with the transverse field $h=U/2$ and the exchange
coupling $J_{ij}=4t\sum_{\sigma}(\langle f_{i\sigma}^{\dag}
f_{j\sigma}^{\phantom{\dag}}\rangle+\mathrm{c.c})$. The transverse Ising
model is a prime example of a system displaying a quantum critical point at
a critical ratio $(2h/J)_c=(U/J)_c$, separating a magnetically ordered phase
from a quantum paramagnet. In the following, we restrict our analysis to
translation-invariant states, for which $J_{ij}= J= -(16/z) \int_{-2td}^0
d\varepsilon\varepsilon\rho_{\sigma}(\varepsilon)\approx 2.67\,t$, and
$\rho_{\sigma}(\varepsilon)$ is the non-interacting density of states
per spin.

\begin{figure}[b]
\includegraphics{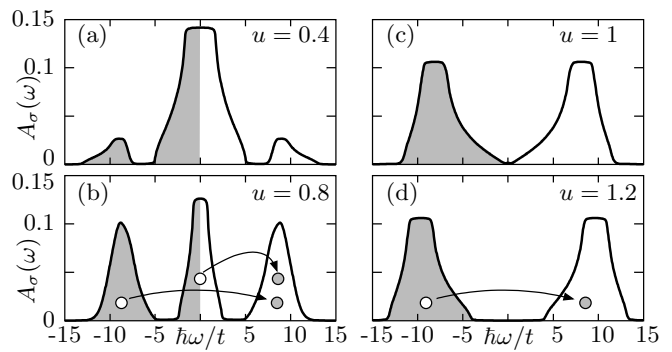}
\caption{
The single-particle spectral density (per spin $\sigma$) $A_{\sigma}(\omega)$
calculated within the slave-spin theory for different interaction strengths
$u$. In the metallic phase, panels (a) and (b), the DGDO probe can induce
transitions to the upper Hubbard band from either the coherent Gutzwiller band
at $\hbar\omega=0$ or the lower Hubbard band, see (b). In the insulating state,
only the latter possibility remains, see (d).
}
\label{fig:spectral}
\end{figure}

Applying a single-site molecular-field approximation to Eq.~(\ref{eq:HTIM}),
the critical ratio is given by $(U/J)_c=z$ and we recover the result of
the Gutzwiller approximation \cite{Gutzwiller63} of the Hubbard model.
Within the molecular-field approximation, the magnetization of the pseudospin
points along the $z$-component of the quantization axis rotated around
the $y$-axis due to the action of the magnetic field $h$ (or interaction
$U$). The angle of rotation is given by $\tan(\phi)=\sqrt{1-u^2}/u$ for
$u\leq 1$ and $\tan(\phi)=0$ for $u>1$, where $u=U/U_{c}$. The fluctuations
around this magnetization are calculated within a spin-wave analysis
\cite{Wang68,Huber07} and we find the gapped pseudospin-wave spectrum
\begin{equation}
\hbar\omega({\bf k})=\frac{U_c}{2}
\times
\left\{\begin{array}{cc}\sqrt{1+\gamma_{\bf k}u^2},
&\quad\mathrm{for}\quad u\leq 1,\\
\sqrt{u^2+\gamma_{\bf k}u},&\quad\mathrm{for}\quad u> 1,\end{array}\right.
\label{eq:omega}
\end{equation}
with $\gamma_{\bf k}=-(1/d)\sum_{i=1}^{d}\cos(k_{i})$ and an excitation
gap $\Delta=\hbar\omega(0)$.

The quantum criticality at $u=1$ is reflected in the softening of the
mode (\ref{eq:omega}). For $u>1$, the jump $\Delta\mu$ in the chemical
potential from hole to particle doping amounts to twice the excitation gap,
$\Delta\mu =2\Delta$; the above pseudospin-mode corresponds to the gapped
charge excitation of the Mott insulator \cite{Lavagna90,Castellani92}.

To estimate the validity of the molecular-field plus spin-wave calculation,
we compare fluctuations with the magnitude of the order parameter and define
$u^{\ssm fl}$ through the condition $1/(z+1)[\sum_{\langle 0,i\rangle}\langle
\delta S_0^x \delta S_i^x\rangle + \langle (\delta S_0^x)^2 \rangle] =
\langle S_0^x\rangle^2|_{u^{\ssm fl}}$. For the cubic lattice, we obtain
$u^{\ssm fl}\approx 0.9$, hence, for $u\approx1\pm0.1$ the fluctuation
induced corrections to the molecular-field result are important and the
validity of the above analysis is limited.

Below, we will see that the DGDO can be qualitatively understood by the
structure of the single-particle spectral density, which is defined
as $A_{\sigma}(\omega)=-\sum_{\bf k} \mathrm{Im}\, G^R_{\sigma}({\bf
k},\omega)/N\pi$. Here, $G^{R}_{\sigma}({\bf k},\omega)$ is the retarded
single-particle Green's function of the $c$-fermions which is calculated in
a straightforward way using the Lehmann representation. The result is shown
in Fig.~\ref{fig:spectral} in the metallic as well as in the insulating
phase.  The gapped mode found in the transverse Ising model leads to the
\emph{incoherent} weight around $\pm \max({U_c,U})/2$ in the spectral
density.  In the metallic phase, we find the characteristic three peak
structure with preformed Hubbard bands centered at $\hbar\omega\approx\pm
U_c/2$ and a coherent Gutzwiller band at $\hbar\omega\approx0$. The
Gutzwiller band disappears at $u=1$ and the Hubbard bands touch at
$\hbar\omega=0$.  For $u\gg 1$ the Hubbard bands assume a constant width
of $U_{c}/2$ and are separated by $U$.  The approximate nature of our
treatment of the spin problem violates $(S_{i}^{x})^{2}\equiv 1/4$ and
the spectral weight fails to be properly normalized. This failure only
manifests itself in the fluctuation regime, however.

The nonlinear DGDO signal is calculated from Eq.~(\ref{eq:Domega}) by
expressing $H_{\ssm int}$ and $K$ in terms of the slave-spin operators;
a careful analysis of the consistency in the order of expansion is
crucial.  For $u\leq1$, we find {\it two} distinct features in $\delta
\mathcal{D}(\omega)$, cf.\ Fig.~\ref{fig:dido-tot}: The sharp peak at
$\hbar\omega=\Delta$ with weight $\delta{\mathcal D}^{<}_{\ssm tot}$
is due to the excitation of a pseudospin-wave mode at ${\bf k} = 0$. In
the fermionic language, this corresponds to the process where the system
is excited from the Gutzwiller band to the upper  Hubbard band, cf.\
Fig.~\ref{fig:spectral}(b). The peak measures the quasiparticle weight $Z$,
which is absent in the insulating state; hence its vanishing serves as a
clear signal for the metal-insulator transition. Without the inclusion of
quasiparticle interactions, the peak is infinitely sharp. The broad peak
around $\hbar\omega\approx \max(U_c,U)$ with weight  $\delta {\mathcal
D}^{>}_{\ssm tot}$ originates from a continuum of pairs of pseudospin waves
and corresponds to the excitation across the gap between the (preformed)
upper and lower Hubbard bands, cf.  Fig.~\ref{fig:spectral}(d); this
feature survives the phase transition and $\delta {\mathcal D}_{\ssm tot}^>$
saturates at $u\gg 1$. The double-peak structure in Fig.~\ref{fig:dido-tot}
is an artifact of our mean-field approach \cite{Huber07}. To obtain the
exact line-shape a more complicated calculation is required.

In conclusion, we have calculated the DGDO in various regimes and
have investigated the suitability of this probe to characterize the
Mott transition in a fermionic system. To make the closest contact to
experiment we find that it is convenient to study the integrated weight
$\delta \mathcal{D}_{\ssm tot}$.

For a characterization of the localized phase our findings show that the
weight $\delta\mathcal{D}^>_{\ssm tot}$ from frequencies $\hbar\omega \approx
U$ is not an optimal measure for Mott physics: All regimes, low density,
high temperatures (single-site problem), as well as the half-filled case
display qualitative similar results. Only a quantitative analysis allows to
distinguish different regimes: the DGDO signal $\delta\mathcal{D}^>_{\ssm
tot}$ in the interaction dominated regime saturates at a \emph{density
dependent} value which is largest at half filling. It turns out, however,
that one can uniquely identify the Mott transition in a strongly frustrated
system by making the distinction between $\delta\mathcal{D}^<_{\ssm tot}$
and $\delta\mathcal{D}^>_{\ssm tot}$. These two weights characterize the
transitions between the Gutzwiller and upper Hubbard and those between
the lower and higher Hubbard bands, respectively. The quantity $\delta
\mathcal{D}^<_{\ssm tot}$ traces the quasiparticle weight $Z$ and captures
the presence of the coherent Gutzwiller band in the metal; its vanishing
then serves as a clear signal to identify the Mott transition. The main
effect of a trapping potential on the disappearance of this coherent signal
is expected to come from a residual signal from the low-density regime in
the periphery of the trap and its influence on the close-by insulator.

We thank the group of T. Esslinger, G. Blatter, M. Sigrist, T.M. Rice, E.
Demler, D. Pekker, R. Sensarma, and L. Pollet for insightful discussions
and acknowledge financial support from the Swiss National Foundation
through the NCCR MaNEP.

\end{document}